\documentclass[pdflatex,sn-nature]{sn-jnl}
\usepackage{graphicx}%
\usepackage{multirow}%
\usepackage{amsmath,amssymb,amsfonts}%
\usepackage{mathrsfs}%
\usepackage[title]{appendix}%
\usepackage{xcolor}%
\usepackage{textcomp}%
\usepackage{manyfoot}%
\usepackage{booktabs}%
\usepackage{algorithm}%
\usepackage{algorithmicx}%
\usepackage{algpseudocode}%
\usepackage{listings}%
\usepackage{graphicx}
\graphicspath{ {./figures/} }
\usepackage[most]{tcolorbox}
\newtcolorbox{llminsight}{boxrule=0.6pt, colback=gray!5, colframe=gray!40, arc=2pt, left=6pt, right=6pt, top=4pt, bottom=4pt}
\theoremstyle{thmstyleone}%
\theoremstyle{thmstyletwo}%
\theoremstyle{thmstylethree}%

\begin{document}

\title[Immunological Density Shapes Recovery Trajectories in Long COVID]{Immunological Density Shapes Recovery Trajectories in Long COVID}


%
%
%
%
%
\author[1,2]{\fnm{Jing} \sur{Wang}}

\author[1]{\fnm{Tong} \sur{Zhang}}

\author[1]{\fnm{Xing} \sur{Niu}}

\author[3]{\fnm{Jie} \sur{Shen}}

\author[4]{\fnm{Yiming} \sur{Luo}}

\author[5]{\fnm{Qiaomin} \sur{Xie}}

\author[6]{\fnm{Amar} \sur{Sra}}

\author[7]{\fnm{Zorina} \sur{Galis}}

\author[1,2]{\fnm{Jeremy C.} \sur{Weiss}}

\affil[1]{\orgname{National Library of Medicine}}

\affil[2]{\orgname{University of Illinois Urbana-Champaign}}

\affil[3]{\orgname{Stevens Institute of Technology}}

\affil[4]{\orgname{Columbia University}}

\affil[5]{\orgname{University of Wisconsin-Madison}}

\affil[6]{\orgname{George Washington University}}
\affil[7]{\orgname{National Heart, Lung, and Blood Institute}}
\abstract{
		Post-acute sequelae of SARS-CoV-2 infection (Long COVID) frequently persists for months, yet drivers of clinical remission remain incompletely defined. 
		Here we analyzed 97,564 longitudinal PASC assessments from 13,511 participants with linked vaccination histories to disentangle passive temporal progression from vaccine-associated change. 
		Using a clinically validated threshold (PASC $\geq 12$), trajectories separated into three phenotypes: Protected (persistently sub-threshold), Refractory (persistently symptomatic), and Responders (transitioning from symptomatic to recovered).
		Across the full cohort, symptom severity increased modestly with elapsed time ($r=0.0521$, $P=1.26\times10^{-59}$), whereas cumulative vaccination showed an inverse association with severity ($r=-0.0434$, $P=5.95\times10^{-42}$).
		Kaplan--Meier analyses further supported distinct recovery kinetics, with Responders recovering gradually over extended follow-up.  
		In summary, baseline Long COVID severity appears clinically deterministic. In the absence of intervention, symptoms typically persist without spontaneous resolution. Recovery is primarily associated with repeated immunization.
}

\keywords{Long COVID, PASC, Vaccination, Immunological Density, Clinical Phenotypes}



\maketitle

\section{Introduction}\label{sec1}

More than 658 million people worldwide have been infected with SARS-CoV-2 (Long COVID) \cite{who}. It is a major and ongoing public health challenge, with substantial impacts on quality of life and health-care utilization \cite{danesh2023symptom,huang2021retracted}. 
Although vaccination reduces risk of acute COVID-19 and incident Long COVID \cite{wang2025active}, its association with symptom resolution among individuals with established PASC remains uncertain, with prior studies limited by small sample size, short follow-up, and heterogeneous outcome definitions. For example, \cite{grady2025impact} has 16 participants and follow up at most 16 weeks. 

In this work, we have 13,511 RECOVER adult cohort study included SARS-CoV-2 infected and uninfected participants with at least 6 month followup visits. Our dataset is discrete time serie dataset, that includes participants' vaccine records, PASC score observation and related timestamp \cite{wang2025large,wang2025mimic}. The PASC score \cite{thaweethai2023development} considers 12 symptoms with corresponding scores ranging from 1 to 8,  such as PEM, fatigue, brain fog, loss of or change in smell or taste, thirst, chronic cough, and so forth. The suggested  threshold is 12.

A central unresolved question is whether PASC improves passively with time or instead requires discrete immunological perturbations to shift clinical state. 
To address this, we convert the participant's trajectories into discrete time series format and let Large Lange Models to read the data first. We propose the prompt to guide the LLM to consider potential noise in the data and using optimization skills in data processing. When feeding LLM with multiple records, our prompt suggest LLM to consider fairness that treats different trajectories equally \cite{wang2025metric}.

With the insights of LLM, we hypothesized that longitudinal courses are heterogeneous and that participants cluster into distinct clinical phenotypes with different sensitivity to vaccination. 
We therefore (i) stratified participants by trajectory into Protected, Responder, and Refractory phenotypes; (ii) constructed temporal and immunological variables capturing days since last vaccination and cumulative dose count; and (iii) tested whether vaccination density, rather than elapsed time, is associated with symptom reduction and state transition. 
This framework enabled us to connect baseline severity, dose exposure, and recovery kinetics within a unified longitudinal analysis.

\section{Results}\label{sec2}
\subsection{Phenotype definition}
\begin{figure}
	\centering
	\label{fig:cohort}
	\includegraphics[width=\textwidth]{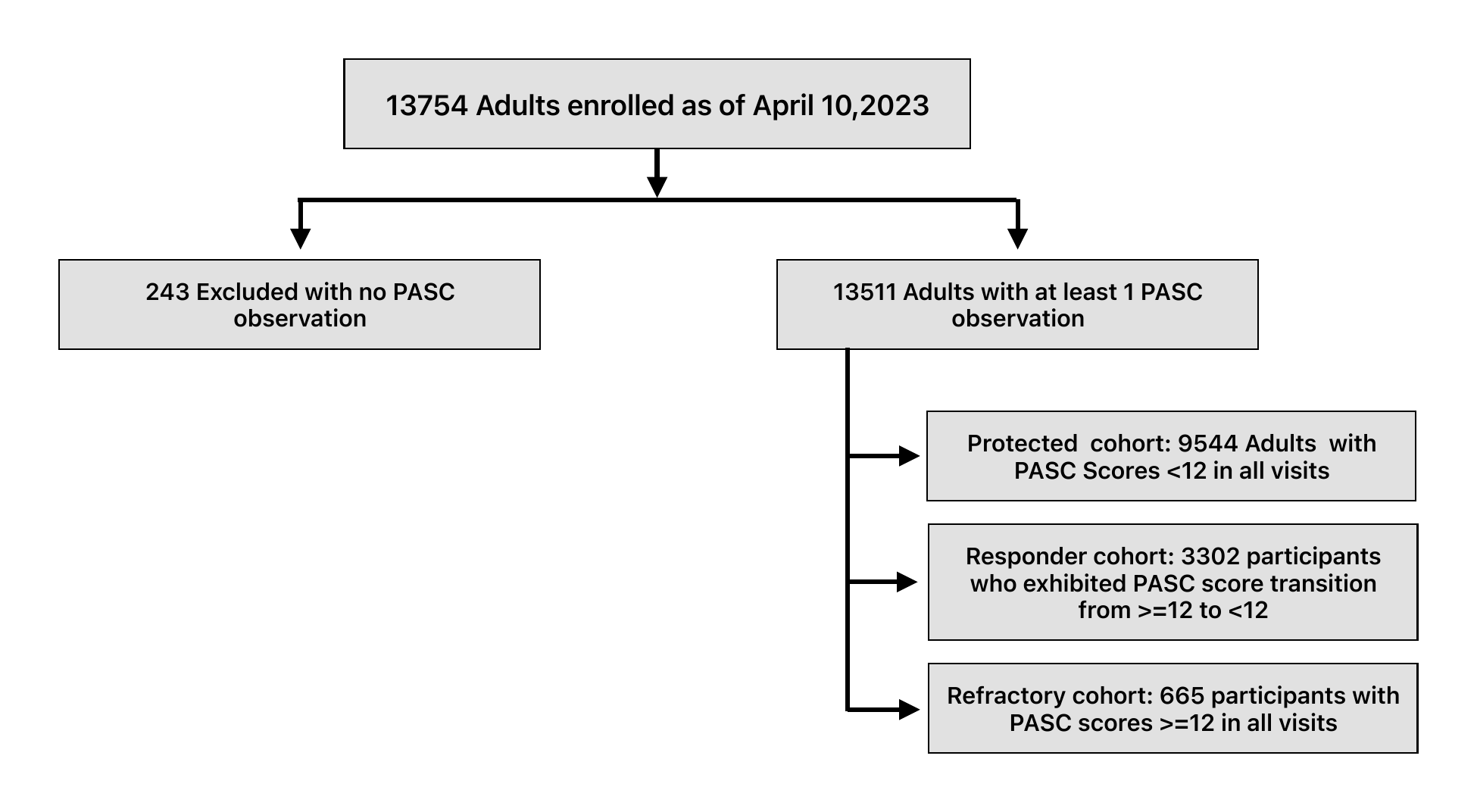}
	\caption{Protected, Responder and Refractory subcohort from RECOVER Adult.}
\end{figure}
Our cohort is from RECOVER Adult \href{https://recovercovid.org/}{dataset}. The RECOVER program recruits participants in the goal to understand, treat, and prevent the postacute sequelae of SARS-CoV-2 infection. There are 13754 participants who were enrolled in the RECOVER adult cohort before April 10, 2023, we select 13511 that completed a symptom survey 6 months or more after acute symptom onset or test date.


We cluster participants clustered into three phenotypes based on longitudinal PASC status as shown in Figure \ref{fig:cohort}: Protected (persistently sub-threshold), Refractory (persistently symptomatic), and Responders (transitioning from symptomatic to sub-threshold). The demographic characteristics are shown in Table \ref{tab:data}.
\begin{table}[t]
	\centering
	\label{tab:data}
	\caption{\textbf{Demographic characteristics.}}
	\begin{tabular}{lccc}
		\hline
		\textbf{Characteristic} & \textbf{Protected} & \textbf{Responder)} & \textbf{Refractory} \\ \hline
		{\#  Participants} &9,544 & 3302 & 665 \\ \hline
		Age at enrollment, median (IQR) & 44.0 (33.0-60) & 49.0 (37-60) & 48.0 (39-57) \\
		\hline
		Age category at enrollment &  &  &  \\
		\quad 18-45 & 3084 & 1365 & 322 \\
		\quad 46-64  & 5014 & 1440 & 290 \\
		\quad $>$65 & 1439 & 495 & 52 \\
		\hline
		Sex assigned at birth &  &  &  \\
		\quad Female & 6637 & 2515 & 545 \\
		\quad Male & 2869 & 783 & 119 \\
		\quad Intersex & 4 & 1 & 0 \\
		\quad Unknown & 33 & 3 & 1 \\
		\hline
	\end{tabular}
\end{table}

\begin{figure}[h!]
	\centering
	\begin{minipage}[t]{0.49\linewidth}
		\centering
		\includegraphics[width=\linewidth]{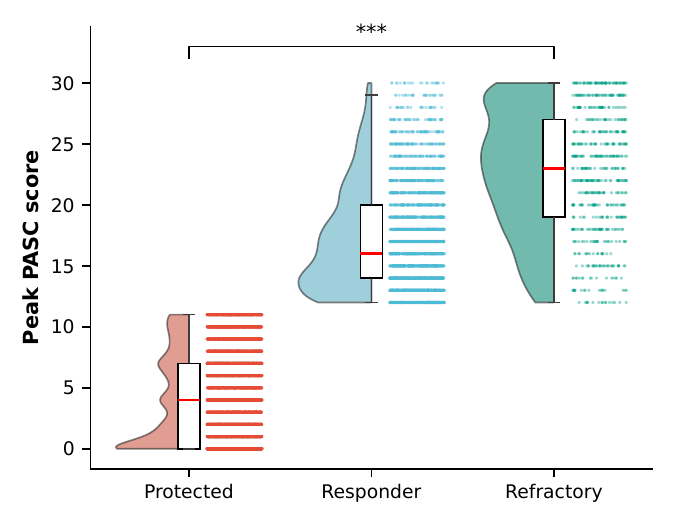}
		\textbf{A}\par
	\end{minipage}\hfill
	\begin{minipage}[t]{0.49\linewidth}
		\centering
		\includegraphics[width=\linewidth]{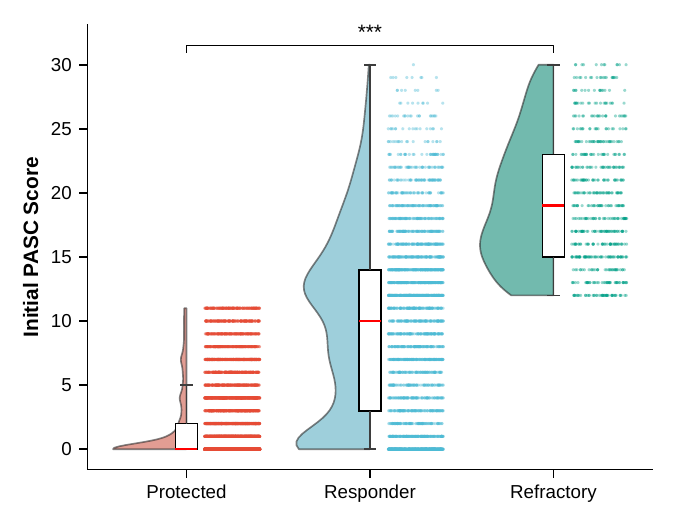}
		\textbf{B}\par
	\end{minipage}
	\caption{\textbf{Distribution of PASC severity across patient cohorts.} \textbf{A,} Distribution of peak PASC severity and individual score variance. \textbf{B,} Distribution of initial PASC severity and individual score variance.}
	\label{fig:peak_initial}
\end{figure}

These phenotypes differed substantially in symptom burden at enrollment (Fig.~	\ref{fig:peak_initial} (A) and at peak severity (Fig.~	\ref{fig:peak_initial} (B)).
Protected participants exhibited low baseline severity and low peak values, Responders showed intermediate baseline severity with a pronounced early peak, and Refractory participants had the highest baseline and peak severity with limited improvement.



%

Figure 	\ref{fig:peak_initial} (A) and (B) plots illustrates the distribution of peak and initial PASC scores across the three identified clinical cohorts: Protected, Responder, and Refractory. Each ``cloud'' represents a half-violin plot showing the kernel density estimation of the scores, while the ``rain'' consists of individual data points jittered to reveal the underlying density of the patient population. The central boxplots indicate the median and interquartile ranges (IQR), highlighting the distinct severity profiles: Protected Cohort demonstrates the lowest peak severity, with a mean score of $4.98 \pm 6.3$. Responder Cohort: Exhibits intermediate severity ($18.40 \pm 7.1$) but shows significant recovery potential post-vaccination. Refractory Cohort: Displays the highest peak PASC scores ($22.80 \pm 5.1$) and persistent symptoms. The significance bracket indicates a highly significant difference between the groups ($p < 0.001$, Kruskal-Wallis $H$-test), confirming that these cohorts represent distinct clinical trajectories of the disease.

\begin{figure}[h!]
	\centering
	\includegraphics[width=\linewidth]{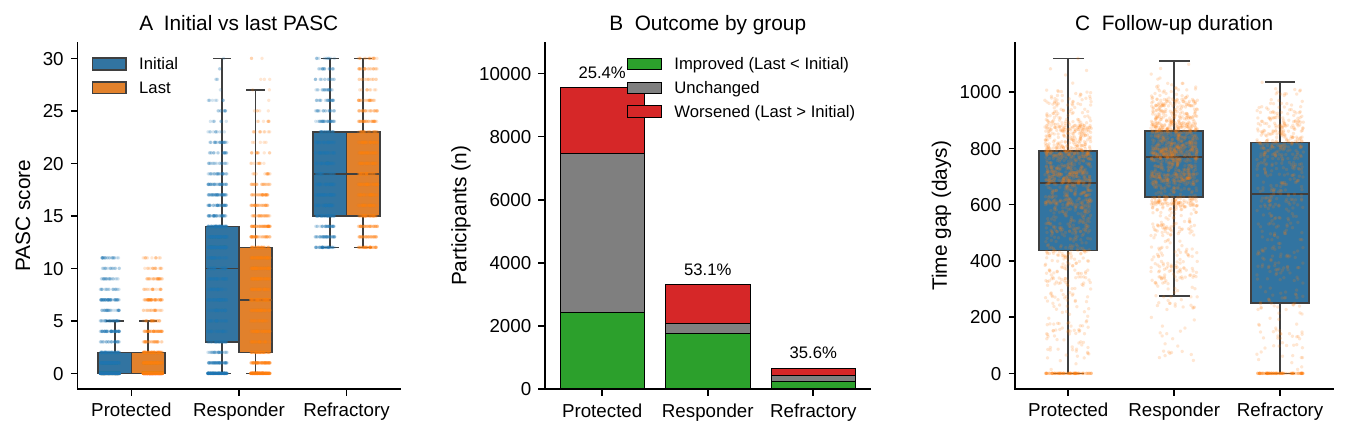}
	\caption{Baseline severity and longitudinal course distinguish Protected, Responder, and Refractory PASC phenotypes.}
	\label{fig:gap}
\end{figure}

Figure \ref{fig:gap} (A) show the distribution of initial and last observed PASC scores for each phenotype. Protected individuals exhibit consistently low symptom burden at both timepoints, Responders show intermediate baseline severity with lower scores at the last visit for many individuals, and Refractory participants maintain the highest symptom burden with limited improvement by the final observation.
Figure \ref{fig:gap} (B) summarizes per-participant longitudinal outcomes based on change from the initial to the last visit (Improved: last < initial; Unchanged; Worsened: last > initial). The Responder cohort contains the largest fraction of individuals showing improvement, whereas the Protected cohort is dominated by unchanged low-burden trajectories and the Refractory cohort shows fewer improvements overall. Percent labels above bars report the proportion improved within each group.
Figure \ref{fig:gap} (C) shows follow-up duration (time gap between first and last recorded PASC score) is comparable across cohorts, supporting that differences in (A-B) reflect distinct clinical trajectories rather than systematic differences in observation window. 

\begin{figure}[h!]
	\centering
	\includegraphics[width=0.75\linewidth]{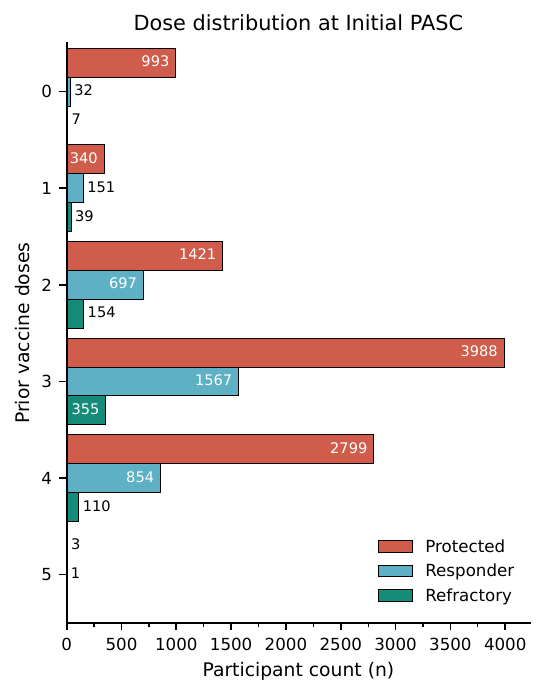}
	\caption{Dose distribution at initial PASC.}
	\label{fig:dose}
\end{figure}

Figure \ref{fig:dose} summarizes the cumulative prior vaccine doses at the time of the earliest observed (index) PASC record. The Protected cohort is concentrated at 3 to 4 prior doses (Dose 3: n=3,988; Dose 4: n=2,799), with fewer participants at 0-2 doses and almost none at $\geq$5 doses. The Responder cohort shows a similar but smaller distribution, again concentrated at 2 to 4 doses (Dose 3: n=1,567; Dose 4: n=854). The Refractory cohort is markedly smaller across all dose strata, peaking at Dose 3 (n=355) with comparatively fewer participants at 0 to 2 doses and very limited representation beyond 4 doses. 


\subsection{Time vs dose}

\begin{figure}[h!]
	\centering
	\includegraphics[width=\linewidth]{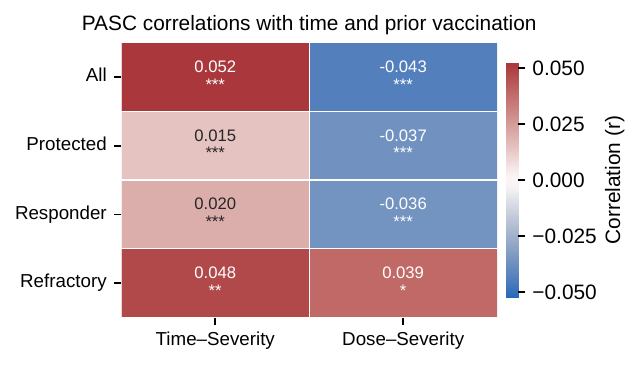}
	\caption{PASC severity shows weak but significant correlations with time and prior vaccination across phenotypes.}
	\label{fig:heatmap}
\end{figure}

To separate passive temporal trends from immunological exposure, we modeled PASC severity against (i) elapsed time since vaccination and (ii) cumulative vaccine dose count. 
Across all observations, severity increased modestly with time ($r=0.0521$, $P=1.26\times10^{-59}$), whereas cumulative vaccination showed an inverse association with severity ($r=-0.0434$, $P=5.95\times10^{-42}$) (Fig.~\ref{fig:heatmap}).

Figure \ref{fig:heatmap} summarizes Pearson correlation coefficients (r) between PASC severity and (i) time (Time-Severity) and (ii) prior vaccine dose count across the full cohort and phenotype-defined subgroups (Protected, Responder, Refractory). Cell color encodes the direction and magnitude of r (diverging scale centered at 0), and each cell is annotated with the corresponding r value and statistical significance. Time-Severity correlations are positive in all groups (All: r=0.052; Protected: 0.015; Responder: 0.020; Refractory: 0.048), indicating slightly higher severity with time. Dose-Severity correlations are negative in All, Protected, and Responder (All: r=-0.043; Protected: -0.037; Responder: -0.036), consistent with a modest reduction in severity with higher prior vaccine dose count, while the Refractory cohort shows a small positive Dose-Severity association (r=0.039). Significance is denoted as $p<0.05 (*)$, $p<0.01 (**)$, and $p<0.001 (***)$.

\subsection{Dose Response}

Our analysis reveals that recovery is not a spontaneous event but is closely tied to cumulative vaccine exposure, particularly within the ``Responder'' phenotype.

%
%
%
\begin{figure}[h!]
	\centering
	\includegraphics[width=0.8\linewidth]{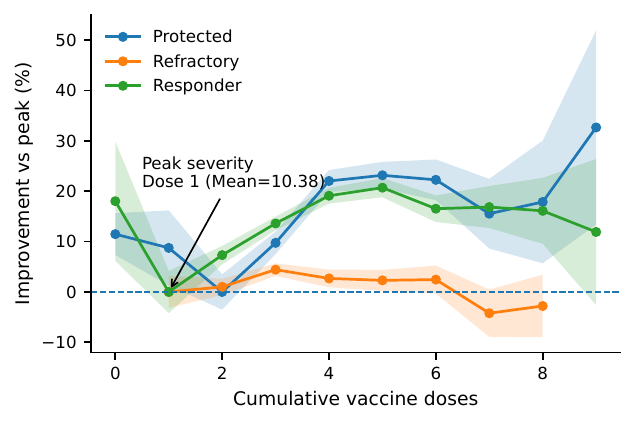}
	\caption{Longitudinal dose-response of PASC severity differs by clinical phenotype.  Mean PASC score ($\pm$95\% CI) is plotted against cumulative prior vaccine doses for Protected, Responder, and Refractory cohorts. The Responder group exhibits a clear post-immunization peak at Dose 1 followed by a predictable recovery, reaching $\sim$20.7\% improvement by Dose 5 relative to the peak. The Protected group maintains consistently low symptom burden with modest additional improvement at higher doses. In contrast, the Refractory group shows little early benefit and a late increase in symptom severity at higher cumulative doses, indicating persistent and worsening clinical burden.}
	\label{fig:reponse}
\end{figure}
Following an initial post-immunization peak at Dose~1 (mean PASC $=10.38$), Responders followed a consistent recovery trajectory with increasing cumulative doses (Fig.~\ref{fig:reponse}). 
From Dose~2 through Dose~5, each additional dose corresponded to an average reduction of $\sim0.54$ PASC points, reaching a 20.71\% improvement by Dose~5 relative to the peak ($P<0.001$; Dose~2: $\approx+7.31\%$; Dose~5: $\approx+20.71\%$). 
The 95\% confidence band indicates stable estimation at common dose strata and widening uncertainty at sparsely populated extremes.

In contrast, the protected cohort exhibited low baseline symptom burden with a shallow dose-response. PASC severity peaked at Dose~2 (mean $=1.79$) and declined through Dose~5 (mean $=1.37$), corresponding to a $23.17\%$ improvement relative to the post-immunization peak (average reduction $\approx 0.14$ points per additional dose from 2 to 5; $P<0.001$). At higher cumulative doses, mean PASC continued to trend downward (Dose~9: mean $=1.20$, $\approx+32.67\%$; Dose~10: mean $=1.08$, $\approx+39.70\%$), although estimates beyond Dose~8 were based on smaller sample sizes.

The refractory cohort showed minimal early improvement and a subsequent worsening pattern. After a modest peak at Dose~1 (mean $=20.54$), PASC scores decreased to a shallow nadir at Dose~3 (mean $=19.63$; $\approx+4.43\%$ improvement; $P=0.011$) but remained near $\sim$20 through Dose~6 ($\approx+2.46\%$). At higher cumulative doses, symptom severity increased markedly (Dose~9: mean $=24.20$, $\approx-17.84\%$; Dose~10: mean $=24.43$, $\approx-18.95\%$ relative to the Dose~1 peak), consistent with persistent and worsening clinical burden.

Although effect sizes were small, the consistency across the full cohort and phenotype-defined subgroups supports an interpretation in which elapsed time alone does not correspond to spontaneous symptom resolution, while cumulative immunological exposure is associated with reduced symptom burden in dose-sensitive phenotypes.

\begin{figure}[h!]
	\centering
	\includegraphics[width=\linewidth]{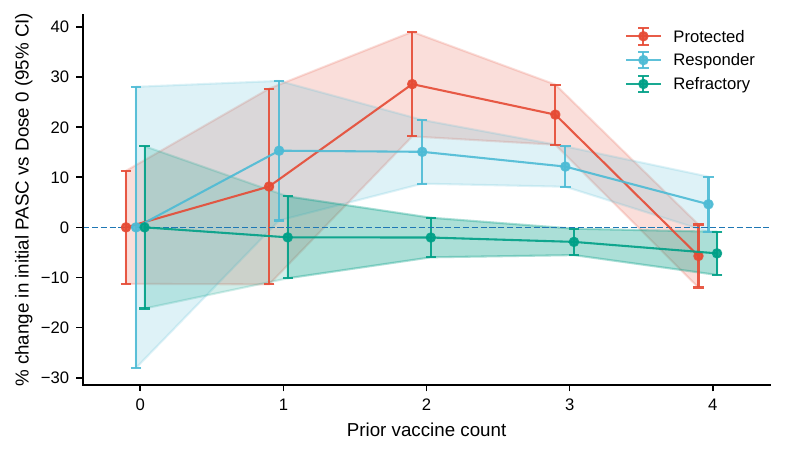}
	\caption{Baseline PASC severity shifts modestly with prior vaccination and differs by phenotype.}
	\label{fig:doseinitial}
\end{figure}

Figure \ref{fig:doseinitial} shows baseline PASC severity shifts modestly with prior vaccination and differs by phenotype. show the percent change in mean initial PASC score relative to Dose 0 across prior vaccine counts (0 to 4) for the Protected, Responder, and Refractory cohorts. Points indicate group-specific means at enrollment, with shaded bands and vertical error bars representing 95\% confidence intervals. The horizontal dashed line denotes no change from Dose 0. Protected and Responder cohorts exhibit modest positive deviations from their Dose-0 baseline at intermediate dose counts, whereas the Refractory cohort remains near baseline with minimal change across doses. Wide uncertainty at Dose 0 in smaller strata reflects limited sample size, emphasizing that observed dose-associated shifts are generally small relative to between-group differences in baseline severity.

\begin{figure}[h!]
	\centering
	\includegraphics[width=0.75\linewidth]{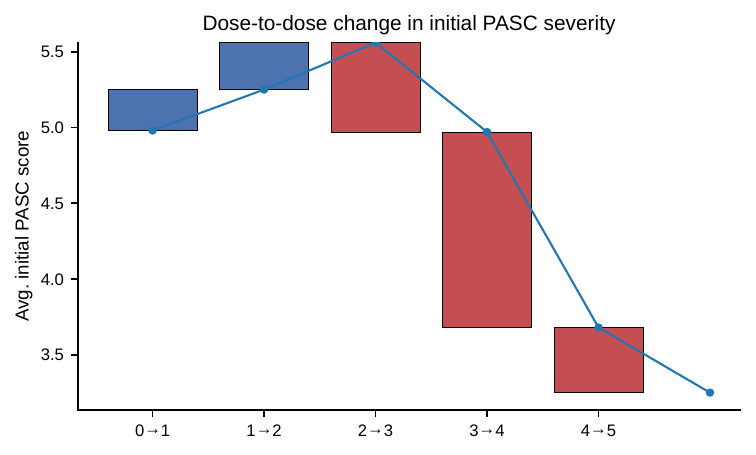}
	\caption{Stepwise shifts in baseline PASC severity across prior vaccine dose strata.}
	\label{fig:overall}
\end{figure}
Figure \ref{fig:overall} plots summarizes of dose-to-dose changes in the mean initial (enrollment) PASC score as prior vaccine count increases from 0 to 5. Each colored block represents the change in mean score between consecutive dose strata, where blue blocks indicate increases in baseline severity and red blocks indicate decreases. The overlaid line traces the absolute mean initial PASC score at each dose level, highlighting a modest rise through Dose 2 followed by a sustained decline, with the largest downward step occurring between Dose 3 and Dose 4.


\subsection{Recovery kinetics}
\begin{figure}[h!]
	\centering
	\includegraphics[width=0.75\linewidth]{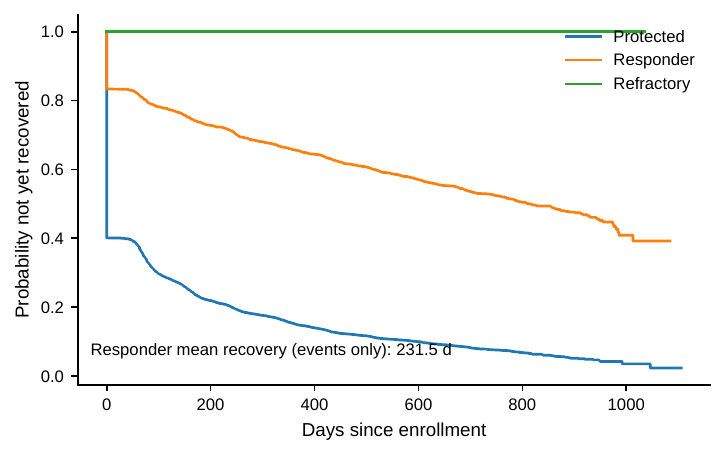}
	\caption{Kaplan-Meier time-to-recovery curves stratified by phenotype.}
	\label{fig:km}
\end{figure}

Kaplan--Meier estimates of time to recovery (first sub-threshold observation) further supported phenotype-specific dynamics (Fig.~\ref{fig:km}). 
Protected participants showed rapid early recovery dynamics, whereas Responders recovered gradually over extended follow-up, consistent with a slow clinical recalibration rather than an acute response. 
Refractory participants remained largely unrecovered throughout observation, consistent with persistent symptom burden.

\section{Methods}\label{sec:methods}

\subsection{Study Design and Cohort Definition}

This longitudinal study utilized 97,564 clinical observations from a cohort of patients monitored for Post-Acute Sequelae of Long Covid. Clinical severity was quantified using a validated PASC scoring metric, where a score $\geq 12$ was defined as symptomatic ``Long COVID'' (Status 1) and a score $< 12$ was defined as recovered or sub-threshold (Status 0). 

\subsection{Temporal and immunological variables}
For each observation, we computed days since the most recent vaccination event ($T_{\mathrm{days}}$) and cumulative vaccine dose count ($D_{\mathrm{count}}$) by chronological summation of recorded doses.

\subsection{Statistical analysis}
Pearson correlation coefficients were used to assess associations between continuous PASC score and (i) time since vaccination and (ii) cumulative dose count, within the full cohort and phenotype strata. 
For binary clinical state analyses, point-biserial correlations were computed where outcome variance was non-zero. 
Time-to-recovery was summarized using Kaplan--Meier estimators, defining recovery as the first observed transition to Status~0; participants without recovery were censored at last follow-up. 
All tests were two-sided with $\alpha=0.05$.

\subsection{Large Language Model-assisted trajectory review}
To qualitatively interrogate longitudinal symptom patterns and generate analysis hypotheses, we performed an LLM-assisted review of participant-level trajectories. Each trajectory consisted of timestamped vaccination events and repeated PASC assessments. We used Qwen3-32B \cite{yang2025qwen3} to summarize trajectories either (i) individually (single participant) or (ii) comparatively (paired participants), using a standardized prompt that explicitly acknowledged potential noise, irregular follow-up intervals, and the observational (non-causal) nature of the records.

In the prompt, we consider noise \cite{shen2021sample,wang2015visual}, fairness \cite{wang2025metric}, feature selection \cite{wang2018provable}, pairwise comparison \cite{shen2021attribute,wang2022fast}, uncertainty \cite{wang2022uncertainty}. That is,

\begin{itemize}
	\item Goal:  describing the participant's course, key events, and strongest observed associations of the longitudal information
	\item Data noise: assume the data may contain noise and artifacts: data missing, vaccine brand typos.
	\item Fairness:  do not infer causality from temporality alone. Consider all the trajectories equaly.
	\item Feature Selection: select most important patterns.
	\item Pairwise comparison:  the summary includes a within-participant comparison (before vs after vaccine events, or early vs late follow-up). Provide at least one pairwise comparison across patients (e.g., one improving vs one persistent), focusing on *patterns*, not identity.
\end{itemize}

Here are the hypothesis from LLM observations.
\begin{llminsight}
	\textbf{LLM insight (hypothesis-generating) based on record of one participant:}
	\emph{Across single-patient summaries, the model frequently highlighted marked within-person variability in PASC scores despite repeated vaccination, and emphasized that the timing of vaccine administrations appeared temporally aligned with short-term changes in severity in some individuals, while other trajectories showed little evidence of improvement.}
\end{llminsight}

\begin{llminsight}
	\textbf{LLM insight (hypothesis-generating) based on records of multiple participants:}
	\emph{In multi-patient comparative summaries, the model repeatedly surfaced three qualitative themes: (1) transient symptom worsening may occur after vaccination in a subset of individuals (consistent with short-lived immune activation); (2) chronic PASC severity is heterogeneous, with a subset exhibiting minimal improvement despite multiple doses; and (3) repeated vaccination was often accompanied by sustained lower PASC scores over longer follow-up in dose-sensitive trajectories, although exceptions were common. These LLM-derived observations were used only for hypothesis generation and to motivate the quantitative design described below.}
\end{llminsight}

\subsection{Longitudinal Records}
\begin{figure}[h!]
	\centering
	\includegraphics[width=0.75\linewidth]{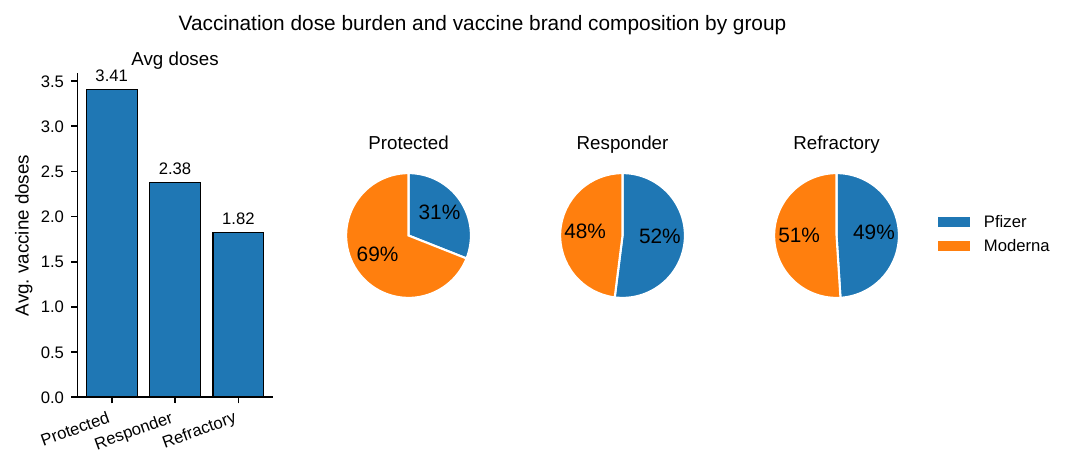}
	\caption{Vaccine dose burden and mRNA vaccine brand composition across Protected, Responder, and Refractory cohorts.}
	\label{fig:pie}
\end{figure}

Each participant record was formalized as a time-stamped event sequence $\{(t_k, e_k)\}_{k=1}^{K}$, where $t_k$ denotes the observation time and $e_k$ denotes the event type. Events included vaccination records (with dose index and brand, e.g., ``Dose~1 Pfizer''), interval immunizations reported between visits (including non-COVID vaccines when present), and PASC assessments recorded at study visits. This representation enabled construction of temporally aligned covariates capturing both passive temporal progression and cumulative immunological exposure. The distribution of vaccine in different brands are as shown in Figure \ref{fig:pie}. 

 Left panel of Figure \ref{fig:pie} shows the mean number of prior vaccine doses at enrollment for each phenotype group (Protected: 3.41; Responder: 2.38; Refractory: 1.82), indicating progressively lower vaccination exposure from Protected to Refractory. Right panels display the distribution of mRNA vaccine brand within each group (Pfizer vs Moderna). Protected participants were predominantly Moderna-vaccinated (31\%/69\%), whereas Responder and Refractory cohorts exhibited a more balanced composition (52\%/48\% and 49\%/51\%, respectively). Together, these plots summarize group-level differences in vaccination intensity and vaccine brand mix at baseline.

\subsection{Temporal and immunological covariates}
Two primary independent variables were defined to separate passive recovery from active intervention. \textbf{Temporal progression} was defined as days since the most recent vaccination event,
\begin{equation}
	T_{\mathrm{days}}(t)= t - \max\{t_k: e_k \in \mathcal{V},\, t_k \le t\},
\end{equation}
where $\mathcal{V}$ denotes vaccination events. \textbf{Immunological density} was defined as the cumulative vaccine dose count at time $t$,
\begin{equation}
	D_{\mathrm{count}}(t)=\sum_{t_k \le t}\mathbb{I}(e_k \in \mathcal{V}_{\mathrm{COVID}}),
\end{equation}
computed by chronological summation of COVID-19 vaccine doses (primary series and boosters).

\subsection{Association models for continuous and binary outcomes}
Let $Y$ denote clinical status derived from PASC assessments. For continuous severity, we used the raw PASC score $Y \in \mathbb{R}$ and computed Pearson product--moment correlations to quantify linear association between $(T_{\mathrm{days}}, D_{\mathrm{count}})$ and severity, preserving sub-threshold and within-state variability.

For binary clinical state, we thresholded PASC using the established criterion (Status~1 if PASC $\ge 12$; Status~0 if PASC $<12$). Associations between continuous covariates and binary state were quantified using the point-biserial correlation,
\begin{equation}
	r_{pb}=\frac{\bar{D}_1-\bar{D}_0}{s_D}\sqrt{\frac{n_1 n_0}{n(n-1)}},
\end{equation}
where $\bar{D}_1$ and $\bar{D}_0$ are mean dose counts among symptomatic and recovered observations, $s_D$ is the standard deviation of $D$, and $n_1,n_0$ are group sizes. Statistical significance was evaluated using the standard $t$-approximation under the null of zero correlation, $t=r\sqrt{(n-2)/(1-r^2)}$. For phenotypes with invariant clinical state (Protected and Refractory), point-biserial estimates are undefined due to zero outcome variance.

\subsection{Survival analysis of time to recovery}
To characterize recovery kinetics, we performed survival analysis using Kaplan--Meier estimators. Recovery time was defined as the first visit at which a participant met the recovery criterion (PASC $<12$) following symptomatic status. Participants without observed recovery were right-censored at their last follow-up time. Kaplan--Meier curves were stratified by phenotype to compare recovery dynamics across Protected, Responder, and Refractory groups.

\subsection{Causal interpretation and limitations}
All analyses are observational. The LLM component served to identify recurring temporal motifs and to motivate the specification of $T_{\mathrm{days}}$ and $D_{\mathrm{count}}$ as primary covariates; it was not used to estimate effect sizes. Correlation and survival analyses quantify association and recovery kinetics but do not by themselves establish causality. Future work will incorporate causal inference approaches (e.g., adjustment for baseline severity and time-varying confounding, or target trial emulation) to more directly estimate vaccine-associated effects on symptom resolution.

\section{Limitations}

Observational Nature: This study is purely observational; while we identify strong associations between vaccination count and symptom reduction, we cannot definitively establish causality or rule out unmeasured confounding factors.

Data Heterogeneity: The reliance on self-reported clinical assessments introduces potential noise, and irregular follow-up intervals may affect the precision of the recovery kinetics observed in the Kaplan-Meier analysis.

Sample Size at Extremes: While the overall cohort is large, data for individuals receiving more than eight vaccine doses are limited, leading to wider confidence intervals and less certain estimates for those specific strata.

LLM Role: The LLM was used exclusively for pattern identification and hypothesis generation; it was not used to calculate formal effect sizes or clinical significance.

\section{Conclusion}
In conclusion, our findings challenge the notion that Long COVID symptoms naturally dissipate over time. Instead, we demonstrate that clinical improvement is significantly associated with cumulative ``immunological density'', repeated vaccine exposure, which acts as a catalyst for state transition in dose-sensitive individuals. The 20.7\% symptom reduction observed in Responders highlights the potential for targeted immunization strategies to drive clinical remission. These results underscore the importance of baseline severity in predicting disease outcomes and suggest that future therapeutic efforts should focus on the underlying mechanisms that render certain phenotypes refractory to standard immunological interventions

\bibliographystyle{sn-nature}

\bibliography{sn-bibliography}

\end{document}